\documentclass[runningheads]{llncs}
\usepackage{graphicx}
\usepackage{color,soul} 
\usepackage{amsxtra}
\usepackage{amscd}
\usepackage{multirow}
\usepackage{hhline}
\usepackage[flushleft]{threeparttable}
\usepackage{cite}
\usepackage{amsmath,amssymb,amsfonts}
\usepackage{algorithmic}
\usepackage{textcomp}
\usepackage{xcolor}
\usepackage{placeins}
\begin{document}
\title{Deep Learning-Aided Spatial Multiplexing with
	Index Modulation}
%

\author{Merve Turhan\inst{1}\orcidID{0000-0002-3150-0687} \and
Ersin \"{O}zt\"{u}rk\inst{2,3}\orcidID{0000-0002-1935-2749} \and
Hakan Ali \c{C}{\i}rpan\inst{3}\orcidID{0000-0002-3591-6567}}
\authorrunning{F. Author et al.}
%
\institute{Ericsson Research, Kista, Sweden \thanks{Work done while author was afﬁliated with Istanbul Technical University} \and Netas, Department of Research and Development  34912, Pendik, Istanbul, Turkey \and 
Istanbul Technical University, Faculty of Electrical and Electronics Engineering 34469, Maslak, Istanbul, Turkey }
\maketitle              
\begin{abstract}
In this  paper, deep  learning  (DL)-aided  data  detection  of  spatial multiplexing (SMX) multiple-input multiple-output (MIMO) transmission with  index  modulation (IM)  (Deep-SMX-IM) has been proposed. Deep-SMX-IM has been constructed by combining a zero-forcing (ZF) detector and DL technique.  The proposed method uses the significant advantages of DL techniques to learn transmission characteristics of the frequency and spatial domains. Furthermore, thanks to using subblock-based detection provided by IM, Deep-SMX-IM is a straightforward method, which eventually reveals reduced complexity. It has been shown that Deep-SMX-IM has significant error performance gains compared to ZF detector without increasing computational complexity for different system configurations.

\end{abstract}
\section{Introduction}

The demand for wireless communications continues to
increase and expand rapidly with new applications. In order to carry out the requested demand, orthogonal frequency division multiplexing (OFDM) has been proposed by Third Generation Partnership Project (3GPP) \cite{3GPP_38201, 3GPP5GNR}. In spite of its proven
advantages, OFDM has some drawbacks such as high out-of band
(OOB) emission and high peak-to-average power ratio
(PAPR) \cite{GWunder}. In this sense, generalized frequency division multiplexing
(GFDM) \cite{Michailow1} came into prominence in terms of reduced latency, high spectral efficiency, and low OOB emission. Also, spatial multiplexing (SMX) is an effective method to improve spectral efficiency. On the other
hand, index modulation (IM) techniques \cite{Ertugrul1} provide spectral and energy efficiency with using transmissions entities to convey digital information innovatively. Taking account of GFDM, MIMO and IM efficiencies, their tight integration has been
introduced in \cite{SMGFDMSuboptimal,GFDMIM,GFDMSFIM,gfdm_fim, gfdm_im_ml_sic,smx_gfdm_im}.
In the past decade, deep neural networks (DNNs) has been widely applied in miscellaneous areas, e.g. speech recognition, object detection, natural language processing \cite{Goodfellow-et-al-2016}. Also, it has become an important area for communication systems, especially for physical layer problems \cite{8054694, Huang_2020}. In \cite{Samuel, MIMODetwDL, ModelDriven_DL_MIMO}, the use of DNN for MIMO detection has been examined. In \cite{DLaidedGFDM,DeepIM, DLaidedGFDM-IM}, deep learning (DL)-aided data detection scheme has been presented for GFDM, OFDM with IM (OFDM-IM) and GFDM with IM (GFDM-IM), respectively.
In this article, a DNN-aided detector is proposed for the combined application of SMX transmission, GFDM, and IM for the purpose of improving error performance without increasing complexity. The main contribution of this article is to adapt a convolutional neural network (CNN) and a fully connected neural network (FCNN) to learn the transmission characteristics of spatial and frequency multiplexing, respectively. Note that, a CNN approach provides a flexible structure for SMX transmission thanks to supporting multi-channel operation and preserves the spatial dependence.  Besides, IM scheme enables to implement
subblock-based and parallel data detection, which simplifies the DL model and
reduces the complexity as well as the processing delay significantly. As far as we know, the proposed method would be the first attempt to implement DL-aided SMX with IM (SMX-IM) detection. It has been shown that the proposed method has a significant bit error rate (BER) gain compared to ZF detector with the same complexity in terms of order of magnitude of complex multiplication (CM).
\section{System Model}
Consider a GFDM-based SMX system with $T$ transmit and $R$ receive antennas. Note that the considered system covers the OFDM-based SMX system. The transmitter gets $PT$ information bits as input. A GFDM symbol, each composing of $M$ subsymbols with $K$ subcarriers, is split into $L$ IM groups. Each group is consisted of $u=MK/L$ subcarrier locations, and $v$ out of $u$ subcarrier locations are activated and used to transmit quadrature amplitude modulation (QAM) symbols. Hence, an IM group carries a $p$-bit binary message $\mathbf{s}_t^l=\left[s_t^l\left(1\right),s_t^l\left(2\right),\ldots,s_t^l\left(p\right)\right]$, for $l=1,\ldots,L$, $t=1,\ldots,T$  and $P=pL$. Each $p$-bit binary message consist of $p_i$ and $p_q$ bits.
While $p_q=v\log_2(Q)$ are mapped by $Q$-array mapper, $p_i=\lfloor\log_2\left(C\left(u,v\right)\right)\rfloor$ are executed to determine sub-carrier locations with reference to lookup table \cite{Ertugrul1}. Therefore, $\alpha = 2^{p_i}$ possible realizations are obtained. Here, $C\left(\mu,\nu\right)$ represents the binomial coefficient and $\lfloor \cdot \rfloor$ is the floor function. As a consequence, IM groups  $\mathbf{d}_t^l=\left[d_t^l\left(1\right),d_t^l\left(2\right),\ldots,d_t^l\left(u\right)\right]^T$,
where $d_t^l\left(\gamma\right) \in  \left\lbrace0,\mathcal{S}\right\rbrace$, is built as using mapping operation with $p$ input bits \cite{GFDMIM}. Here, $\mathcal{S}$ denotes $Q$-ary constellation. The resulting IM groups are then concatenated to form the GFDM-IM symbol 
\begin{equation}
\mathbf{d_{t}} = \left[d_{t,0,0},\ldots,d_{t,K-1,0},d_{t,0,1},\ldots,d_{t,K-1,1}\right.\left. ,\ldots,d_{t,K-1,M-1}\right]
\label{eq:gfdm_xad3}
\end{equation}
where $d_{t,k,m}\in  \left\lbrace0,\mathcal{S}\right\rbrace$, for $m=0,\ldots,M-1$, $k=0,\ldots,K-1, t=1,\ldots,T$, is the data symbol of $k$-th subcarrier on $m$-th subsymbol of a GFDM symbol belonging to $t$-th transmit antenna. Then, the GFDM-IM symbol $\mathbf{d}_t$ is modulated by a GFDM modulator and the resulting GFDM transmit signal can be written in linear form as
\begin{equation}
\mathbf{x_t}=\mathbf{A}{\mathbf{d_t}},
\label{eq:gfdm_xad}
\end{equation}	
where $\mathbf{A}$ represents an $MK \times MK$ GFDM transmitter matrix \cite{Michailow1}. Eventually, a cyclic prefix (CP) with length $N_{\text{CP}}$ is attached to $\mathbf{x_t}$ and the resulting vector $\tilde{\mathbf{x_t}}=\left[\mathbf{x_t}\left(MK-N_{\text{CP}}+1:MK\right)^T, \mathbf{x_t}^T\right]^T$
is transmitted over a frequency-selective channel.

At the receiver side, under the assumption that CP is longer than the maximum delay spread of the channel $(N_{\textrm{Ch}})$, the whole received signal can be obtained as
\begin{align}
\underbrace{\begin{bmatrix}\mathbf{y}_{1}\\ \vdots \\ \mathbf{y}_{R} \end{bmatrix}}_{\mathbf{y}}&= \underbrace{\begin{bmatrix}\mathbf{H}_{1,1}\mathbf{A} & \ldots & \mathbf{H}_{T,1}\mathbf{A} \\  \vdots & \ddots & \vdots\\ \mathbf{H}_{R,1}\mathbf{A}& \ldots & \mathbf{H}_{R,T}\mathbf{A} \end{bmatrix}}_{\widetilde{\mathbf{H}}}& \underbrace{\begin{bmatrix}\mathbf{d}_{1}\\ \vdots \\ \mathbf{d}_{T} \end{bmatrix}}_\mathbf{d}& +\underbrace{\mathbf{\begin{bmatrix}\mathbf{n}_{1}\\ \vdots \\ \mathbf{n}_{R} \end{bmatrix}}}_{\mathbf{n}}&
\label{eq:gfdm_basic_system_model}
\end{align}
after the removal of CP. Here, $\mathbf{y}_r=[y_r(0), y_r(1),\ldots,y_r(N-1)]^{\text{T}}$ is the vector of the received signals, ${\mathbf{H}_{r,t}}$, for $t=1,\ldots,T, r=1,\ldots,R$, indicates the $N \times N$ circular convolution matrix formed from the channel impulse response coefficients given by $\mathbf{h}_{r,t}=\left[h_{r,t}(1),h_{r,t}(2),\ldots,h_{r,t}(N_{\textrm{Ch}})\right]^\text{T}$, and $\mathbf{n}_r$ denotes an $N \times 1$ additive white Gaussian noise (AWGN) vector. The elements of $\mathbf{h}_{r,t}$ and $\mathbf{n}_r$ follow $\mathcal{CN}(0,1)$ and $\mathcal{CN}(0,\sigma_n^2)$ distributions, respectively. Here, $\mathcal{CN}(\mu,\sigma^2)$ represents the distribution of a circularly symmetric complex Gaussian random variable with mean $\mu$ and variance $\sigma^2$. Eq. \ref{eq:gfdm_basic_system_model} can be rewritten in a more succinct form as
\begin{equation}
\mathbf{y}=\widetilde{\mathbf{H}}\mathbf{d}+\mathbf{n}.
\label{eq:gfdm_equivalent_system_model}
\end{equation}

\begin{figure*}[!t]
	\centering
	\begin{center}{\includegraphics[scale=0.45]{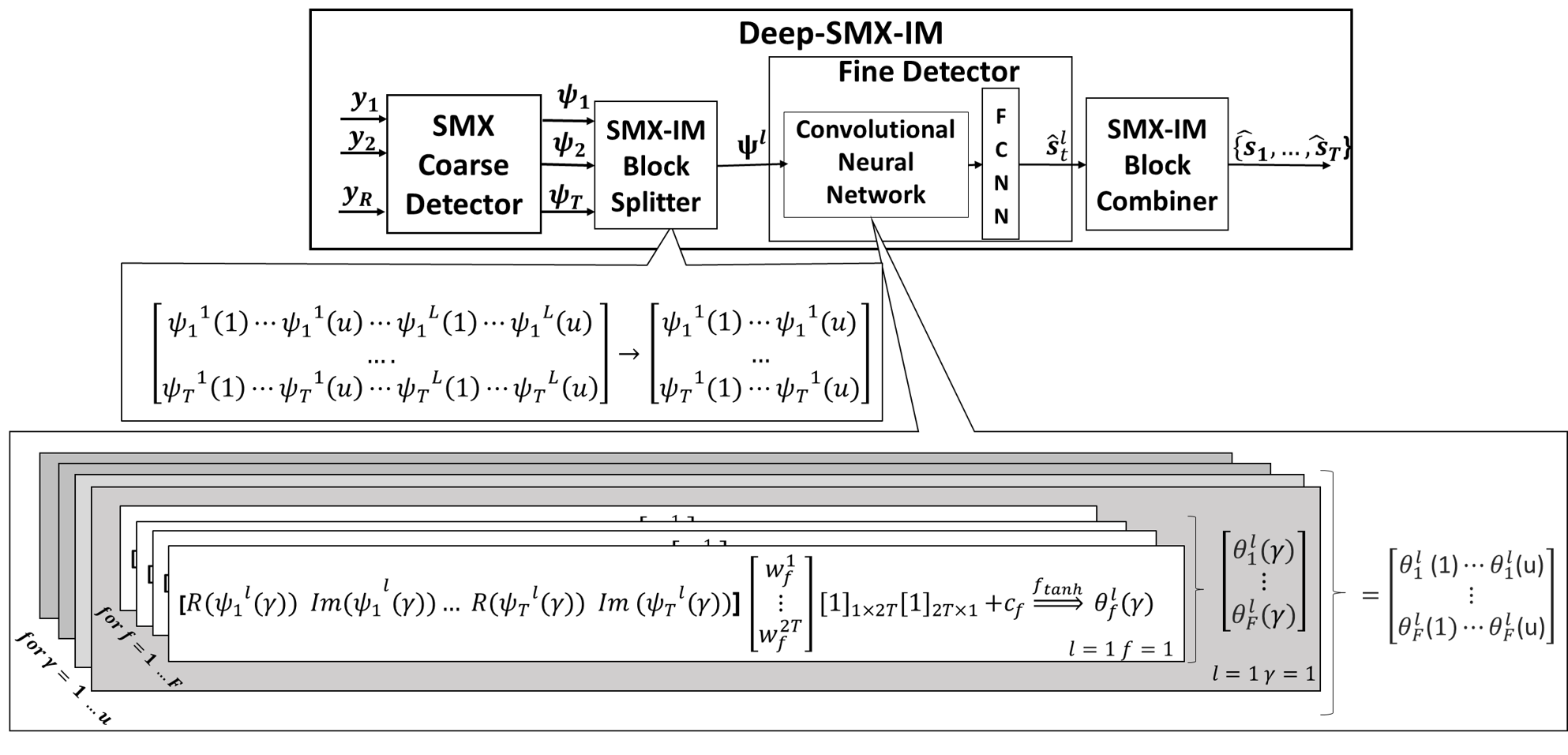}}
		\centering
		\caption{Block diagram of the Deep-SMX-IM.}
		\label{fig:dl_jdd}
	\end{center}
\end{figure*}

\section{Deep Detector}
The block diagram of the proposed deep learning-aided data detection of spatial multiplexing  scheme, termed as Deep-SMX-IM, is given in Fig \ref{fig:dl_jdd}. The channel information is assumed to be perfectly known at the receiver. The proposed detector is built as two stages, namely coarse detector and fine detector. It also has two intermediate steps for regularizing coarse detector output and fine detector outputs, which are SMX-IM Block Splitter and Combiner, respectively. As a first stage, coarse detector applies ZF detection to handle channel and GFDM modulation effects together and the coarse detector's output can be given by
\begin{align}
\begin{bmatrix}\boldsymbol{\psi}_{1}\\ \boldsymbol{\psi}_{2}\\  \vdots \\ \boldsymbol{\psi}_{T} \end{bmatrix}&=\left(\widetilde{\mathbf{H}}^{H}\widetilde{\mathbf{H}}\right)^{-1}\widetilde{\mathbf{H}}^{H}\mathbf{y}{\color{blue},}
\end{align}
where $\boldsymbol{\psi}_{t} = [\boldsymbol{\psi}^{1^T}_{t},\boldsymbol{ \psi}^{2^T}_{t} ,...,\boldsymbol{\psi}^{L^T}_{t}]^T$, for $t = 1,...T$ and for $l = 1,...L$, $\boldsymbol{\psi}^{l}_{t}$ denotes a $u\times1$ vector. After that, coarse detector output $\boldsymbol{\psi}^{l}_{t}$ is subdivided into IM subblocks by SMX-IM Block Splitter and the resulting matrix can be expressed as
\begin{align}
\mathbf{\Psi}^l&= \begin{bmatrix}\psi_1^l(1) &\ldots & \psi_1^l(u) \\  \vdots & \ddots & \vdots\\ \psi_T^l(1)& \ldots & \psi_T^l(u) \end{bmatrix}
\label{eq:psi}.
\end{align}
The fine detector stage of Deep-SMX-IM is built by using CNN and FCNN. The fine detector's CNN part convolves the IM subblock $\mathbf{\Psi}^l$ with the kernels $\bold{w}^{f} = \big[w_{1}^f,\ldots, w_{2T}^{f}\big]${\color{black},} adds bias ${\bold{c}_{f}}$ for $f = 1,...F$ and stride $1$, and the modified received IM subblock is obtained as
\begin{eqnarray}\nonumber
{\theta^l_f}(\gamma) = f_{tanh}(\Re(\psi_{1}^l(\gamma)*w^{f}_1)+ \Im(\psi_{1}^l(\gamma)) *w_{2}^f+ \ldots \\ + \Re(\psi_{T}^l(\gamma)*w_{2T-1}^f)+\Im(\psi_{T}^l(\gamma)) *w_{2T}^f+c_f)
\end{eqnarray}
for $f = 1,...,F$ and $\gamma = 1,...,u$. Note that since complex numbers are not supported by any DL framework yet, real and imaginary parts of the received signals are processed separately. The CNN part repeats the convolution operation for $F$ different kernel filters. The output from CNN can be obtained as
\begin{align}
\mathbf{\Theta}^l&= \begin{bmatrix}\mathbf{\theta}_1^l(1) &\mathbf{\theta}_1^l(2) \ldots & \mathbf{\theta}_1^l(u) \\  \vdots & \ddots & \vdots\\ \mathbf{\theta}_F^l(1)& \ldots & \mathbf{\theta}_F^l(u) \end{bmatrix}.
\label{eq:theta}
\end{align}
After that, $\mathbf{\Theta}^l$ is converted into a vector by flattening process and the resulting vector can be expressed as $\bold{\theta}^l = [ \mathbf{\theta}_1^l(1),\ldots, \mathbf{\theta}_1^l(u),\ldots,\mathbf{\theta}_F^l(1),\ldots,\mathbf{\theta}_F^l(u) ]^T$. The fine detector's FCNN part executes $\mathbf{\theta}^l$ with using {\color{black} $\left\{\mathbf{a,b}\right\}$ parameters, where $\mathbf{a} = {\big[\mathbf{a}_{1} ,\mathbf{a}_{2}]}
	$ includes weight parameters and $\mathbf{b} = [{b}_{1}, {b}_{2}]$ contains bias parameters.} This part consists of only two layers, first layer and output layer have $\tau$ and $PT$ neurons, respectively. Fine detector's output is obtained as 
\begin{equation}
\bold{\hat{s}}^l = f_{sigmoid}(\mathbf{a}_{2}({{f_{tanh}}(\mathbf{a}_{1}\mathbf{\theta}^l+b_{1})})+b_{2},
\label{eq:8}
\end{equation}
where $f_{sigmoid}, f_{tanh}$ are activation functions. Finally, SMX-IM Block Combiner composes the fine detector's output into transmitted information bits.
Before  using  the  proposed  Deep-SMX-IM  detector, it has to be trained offline  with  the  data to be generated at training signal-to-noise ratio (SNR) value by simulations. While the training step uses fixed SNR value, the testing step uses a range of SNR values. Deciding training SNR value has a key role against overfitting. The training executes on total trainable parameters, which consist of $\bold{w}^{f},\bold{c}^{f},\mathbf{a},\bold{b}$, for the purpose of minimizing the loss function. It can be expressed as $loss(\bold{s}^l,\bold{\hat{s}}^l)$ = $\|\bold{s}^l-\bold{\hat{s}}^l\|$. In the training stage, total trainable parameters are randomly initialized at first. Throughout the training, stochastic gradient descent algorithm executes on these parameters, it can be expressed as 
\begin{equation}
{\bold{\epsilon_{+}} = \bold{\epsilon} - \eta\bigtriangledown loss(\bold{\mathbf{s}}^l_B,\hat{\bold{s}}^l_B)},
\end{equation}
where $\bold{\epsilon}$, $\eta$, $B$, represent total trainable parameters, learning rate and batch size respectively.
\section{Complexity Analysis for Deep-SMX-IM} \label{complexity_analysis}
In this section, we have assessed the computational complexity of ZF, maximum likelihood (ML) and Deep-SMX-IM detectors in terms of number of CM. The results are provided in Table \ref{tab:compucomp_smx_details}. Here, $\Psi_{J\times I}$ and $\Phi_{J\times I}$ stand for $J\times I$ matrices, $\psi_{J\times 1}$ and $\phi_{J\times 1}$ denotes $J\times 1$ vectors. Note that, Deep-SMX-IM has only real operation and one CM can be executed with three real multiplications at least. So, the number of multiplications of Deep-SMX-IM is divided by three for the purpose of expressing them as CM.  As seen in Table \ref{tab:compucomp_all}, while ML detector has the highest complexity, Deep-SMX-IM and ZF have approximately the same complexity in terms of order of magnitude of CMs.
\begin{table*}[!t]
	\fontsize{7.5}{8.2}\selectfont
	\begin{center}
		\begin{threeparttable}
			\caption{Comparison of Computational Complexities}
			\label{tab:compucomp_smx_details}
			\begin{tabular}[c]{|l||c|c|c|c|} \hline
				\textit{Receiver Scheme} & \textit{Process} & \textit{Operation} & \textit{Execution Count} & \textit{CMs}  \\ \hline \hline				
				\multirow{2}{*}{ML} & Forming $\widetilde{\mathbf{H}}$ & ${\Phi_{N\times N}\Psi_{N\times N}}^{\dagger}$  & 1 & $N_\text{Ch}N^2RT$ \\ \hhline{~----}
				& Decision & $\text{min}\left({\lVert\phi_{N\times 1}-\left(\Phi_{N\times N}\psi_{N\times1}\right)\rVert}^2\right)^{\dagger\dagger}$ & $\left(\alpha Q^v\right)^{TN/u}$ &  $\left(\alpha Q^v\right)^{TN/u}\left((N^2TRv)/u+NT\right)$ \\ \hline
				\multirow{3}{*}{ZF} & Forming $\widetilde{\mathbf{H}}$ & ${\Phi_{N\times N}\Psi_{N\times N}}^{\dagger}$  & RT & $N_\text{Ch}N^2RT$ \\ \hhline{~----}
				& JDD & ${\left({\Phi_{NR\times NT}}^\text{H}{\Phi_{NR\times NT}}\right)}^{-1}{\Phi_{NR\times NT}}^\text{H}\phi_{NR\times 1}$ & 1 &  $2N^3T^2R+N^3T^3+N^2RT$ \\ \hhline{~----}
				& Decision & $\text{min}\left({\lVert\phi_{u\times 1}-\psi_{u\times1})\rVert}^2\right)$ & $(N/u)\alpha Q^vT$ &  $N\alpha Q^vT$ \\ \hline
				\multirow{4}{*}{Deep-SMX-IM} & Forming $\widetilde{\mathbf{H}}$ & ${\Phi_{N\times N}\Psi_{N\times N}}^{\dagger}$  & RT & $N_\text{Ch}N^2RT$ \\ \hhline{~----}
				& JDD & ${\left({\Phi_{NR\times NT}}^\text{H}{\Phi_{NR\times NT}}\right)}^{-1}{\Phi_{NR\times NT}}^\text{H}\phi_{NR\times 1}$ & 1 &  $2N^3T^2R+N^3T^3+N^2RT$ \\ \hhline{~----}
				& CNN & ${\left(2FT+T\lambda\right)/3}^{\dagger\dagger\dagger}$ & N & $\left(2FT+F\lambda)\right)N/3$ \\ \hhline{~----}
				& FCNN & ${\left(uF\tau+\tau\lambda+\tau pT+pT\delta\right)/3}^{\dagger\dagger\dagger}$ & L & $\left(uT\tau+\tau\lambda+\tau pT+p\delta T\right)L/3$\\ \hline		
			\end{tabular}
			\begin{tablenotes}
				\small
				\item $^{\dagger}$ In every row of ${\mathbf{H}}$, only $N_{Ch}$ out of $N$ elements are non-zero.
				\item $^{\dagger\dagger}$ In $\psi$, only $vML$ complex elements are nonzero.
				\item $^{\dagger\dagger\dagger}$ $\lambda$ and $\delta$  denote to number of real multiplications needed for $f_{tanh}$ and $f_{sigmoid}$, respectively.
			\end{tablenotes}
		\end{threeparttable}	
	\end{center}
\end{table*}

\begin{table*}[!t]
	\begin{center} 
		\fontsize{7.5}{8.2}
		\caption{The Total Number of CMs}
		\label{tab:compucomp_all}
		\begin{tabular}[c]{|l||c|c|c|c|} \hline
			\textit{Config. (T,R)} & \textit{ML} & \textit{ZF} & \textit{Deep-SMX-IM}   \\ \hline \hline
			BPSK  $(2,2)$ & $3.29067\times10^{76}$ & $4.53561\times10^{8}$ &  $4.53282\times10^{8}$ \\ \hline
			4-QAM $(2,2)$ & $ 3.08764 \times10^{134}$  & $4.53840\times10^{8}$ & $4.53328\times10^{8}$\\ \hline
			BPSK  $(4,4)$& $ 1.83301\times10^{178}$  &  $2.31930\times10^{11}$ &  $2.31929\times10^{11}$ \\
			\hline
			4-QAM, $(4,4)$ & $1.08149\times10^{294}$  & $2.31931\times10^{11} $ & $2.31929\times10^{11}$  \\
			\hline
		\end{tabular}
	\end{center}
\end{table*}
\section{Simulation Results and Discussion}
\begin{figure}[!h]
	\centering
	\begin{center}{\includegraphics[scale = 0.40]{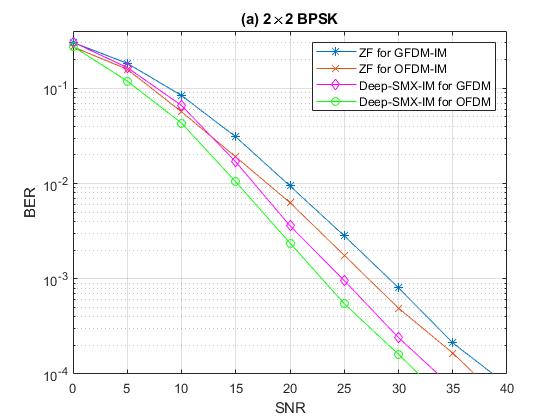}}
		\caption{Performance comparison of Deep-SMX-IM and ZF for $2 \times 2$ SMX-IM using BPSK transmission}
		\label{fig:bpsk}
	\end{center}
\end{figure}
\begin{figure}[!h]
	\centering
	\begin{center}{\includegraphics[scale = 0.40]{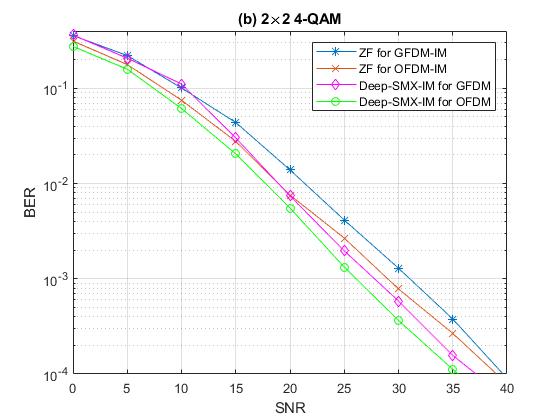}}
		\caption{Performance comparison of Deep-SMX-IM and ZF for $2 \times 2$ SMX-IM using 4-QAM transmission.}
		\label{fig:224qam}
	\end{center}
\end{figure}
\begin{figure}[!h]
	\centering
	\begin{center}{\includegraphics[scale = 0.4]{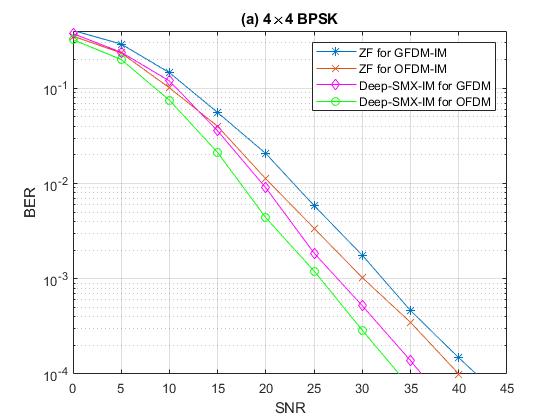}}
		\caption{Performance comparison of Deep-SMX-IM and ZF for $4 \times 4$ SMX-IM using BPSK transmission.}
		\label{fig:44bpsk}
	\end{center}
\end{figure}
\begin{figure}[!h]
	\centering
	\begin{center}{\includegraphics[scale = 0.4]{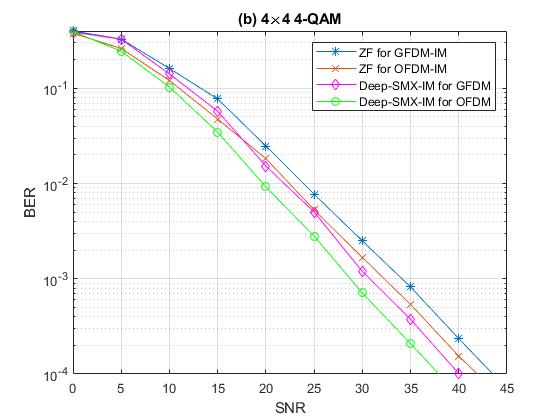}}
		\caption{Performance comparison of Deep-SMX-IM and ZF for $4 \times 4$ SMX-IM using 4-QAM transmission.}
		\label{fig:444qam}
	\end{center}
\end{figure}
\begin{table}[!t]
	\begin{center}
		\fontsize{7.5}{8.2}
		\caption{Deep-SMX-IM Model Parameters}
		\label{tab:model_params}
		\begin{tabular}{|c|c| c|c|c|} 
			\hline
			\textit{Antenna Configuration} & \textit{Modulation}& \textit{Parameter} & \textit{Value}  \\ \hline \hline  
			$2\times2$ & BPSK & $F$ & 64  \\ 
			& & $\tau$ & 128 \\
			& 4QAM & $F$ & 64  
			\\
			&  & $\tau$ & 256 
			\\ \hline
			$4\times4$ & BPSK & $F$ & 128  \\ 
			& & $\tau$ & 256 \\
			& 4QAM & $F$ & 128  \\
			&  & $\tau$ & 512 \\ \hline
		\end{tabular}
	\end{center}
\end{table}
\begin{table}[!t]
	\begin{center} 
		\caption{Fine Detector  Model Summary}
		\label{tab:model_summary-}
		\begin{tabular}{|c|c|c|}
			\hline
			\textit{Layer} & \textit{Output Shape} & \textit{Activation Func.} \\	
			\hline	
			\hline Input ($\mathbf{\Psi}^l$) & ($B$,$u$,$2$,$T$) & None \\ 
			\hline CNN & ($B$,$1$,$u$,$F$) & $f_{tanh}$ \\ 
			\hline Flattening & ($B$,$uF$) & None \\
			\hline 1. Layer of FCNN & ($B$,$\tau$) & $f_{tanh}$ \\
			\hline 2. Layer of FCNN & ($B$,$pT$) & $f_{sigmoid}$ \\
			\hline
		\end{tabular}
	\end{center}
\end{table}
In this section, the performance of Deep-SMX-IM detector has been evaluated for Rayleigh fading with Extended Pedestrian A (EPA) channel model \cite{3GPP_epa} employing BPSK and 4-QAM modulation. The raised cosine filter is used as a GFDM prototype filter with a roll-off factor of 0.5. The following GFDM-IM parameters are assumed: $K = 32, M = 3, N_{Ch} = 8, u = 4, v = 2$. In order to select
the active subcarrier indices, the lookup
table \ref{tab:Look_up1} is used. Note that $K = 32, M = 1$ is assumed for OFDM parameters. Training data including $12\times10^5$ IM groups, is generated at SNR $15$dB according to GFDM and OFDM parameters. The Deep-SMX-IM model is trained 120 epochs with $B = 1000$. In Table \ref{tab:model_params} and \ref{tab:model_summary-}, Deep-SMX-IM fine detector model parameters and summary can be seen respectively. In order to find a global minimum, stochastic gradient based Adam optimizer \cite{Adam}, is used with $8\times10^{-4}$ learning rate.
Fig. \ref{fig:bpsk} depicts the BER comparison of the Deep-SMX-IM and ZF detectors using BPSK along with SMX-GFDM-IM and SMX-OFDM-IM for $2 \times 2$ SMX systems. As seen from Fig. \ref{fig:bpsk}, Deep-SMX-IM provides $5.5$ dB better BER performance than ZF for BPSK.
Fig. \ref{fig:224qam} depicts the BER comparison of the Deep-SMX-IM and ZF detectors using $4$-QAM along with SMX-GFDM-IM and SMX-OFDM-IM for $2 \times 2$ SMX systems. From Fig. \ref{fig:224qam} for a BER value of $10^{-4}$, it is observed that the Deep-SMX-IM for GFDM and OFDM achieves $3$ dB and $4$ dB better BER performance than ZF for SMX-GFDM-IM and SMX-OFDM-IM, respectively.
Fig. \ref{fig:44bpsk} compares the BER performance of the Deep-SMX-IM and ZF employing BPSK along with SMX-GFDM-IM and SMX-OFDM-IM for $4 \times 4$ SMX transmission. As seen from Fig. \ref{fig:44bpsk}, the Deep-SMX-IM for GFDM and OFDM achieves $5.5$ better BER performance than ZF for BPSK at a BER value of $10^{-4}$ for SMX-GFDM-IM and SMX-OFDM-IM, respectively.
Fig. \ref{fig:444qam} compares the BER performance of the Deep-SMX-IM and ZF employing $4$-QAM along with SMX-GFDM-IM and SMX-OFDM-IM for $4 \times 4$ SMX transmission. As seen from Fig. \ref{fig:44bpsk}, Deep-SMX-IM for GFDM and OFDM achieves $3$ dB better BER performance than ZF for $4$-QAM at BER value $10^{-4}$.
As seen from Fig.  \ref{fig:bpsk}  and  \ref{fig:44bpsk}, as spectral efficiency and the modulation order increases, model's learning capacity decreases. However, Deep-SMX-IM continues to retain its advantage over the classical linear detector in all conceivable system parameters. In Table \ref{tab:compucomp_all}, the number of CMs required for Fig. \ref{fig:bpsk} and \ref{fig:44bpsk} are provided. Notice that, while DeepConvIM in \cite{DLaidedGFDM-IM} has an intermediate solution for GFDM-IM, Deep-SMX-IM can be assessed as an efficient solution in terms of computational complexity. 
\begin{table}[!t]
	\begin{center}
		\caption{A look-up table example for $u=4,v=2$.}
		\label{tab:Look_up1}
		\begin{tabular}[c]{|c||c||c|} \hline
			\textit{Bits} & \textit{Indices} & \textit{IM block}  \\ \hline \hline
			$[0\,\,\, 0]$ & $\left\lbrace 1, 2\right\rbrace $ & $\begin{bmatrix}s_{\chi} & s_{\zeta} & 0 & 0   \end{bmatrix}^T$  \\ \hline
			$[0\,\,\, 1]$ & $\left\lbrace 2,3\right\rbrace$ & $\begin{bmatrix}0 & s_{\chi} & s_{\zeta} & 0   \end{bmatrix}^T$ \\ \hline
			$[1\,\,\, 0]$ & $\left\lbrace 3, 4\right\rbrace$ & $\begin{bmatrix} 0 & 0 & s_{\chi} & s_{\zeta}     \end{bmatrix}^T$ \\ \hline
			$[1\,\,\, 1]$ & $\left\lbrace 1,4\right\rbrace$ & $\begin{bmatrix}s_{\chi} & 0 & 0 &  s_{\zeta}   \end{bmatrix}^T$  \\ \hline
		\end{tabular}
	\end{center}
\end{table}

\section{Conclusion}
A novel DL-aided detector, called Deep-SMX-IM, has been proposed for SMX transmission with IM. It has been shown that the proposed deep learning-aided detector provides important BER performance improvement compare to ZF detector with the approximately same complexity in terms of order of magnitude of CMs. Our results highlight that the significant advantages of deep learning techniques should be engineered to overcome the challenges of wireless communications arising from the distinct characteristics of time, frequency and spatial domains. As future work, performance analysis of Deep-SMX-IM can be examined for the different coarse detectors, e.g. minimum mean-squared error (MMSE), ML. Furthermore, it may be considered to apply the proposed methods in different index modulation concepts.
%
%
%
\bibliographystyle{splncs04}
\bibliography{smx_gfdm_im}
\end{document}